# Re-discovering micro-emulsion electrolytes: a biphasic electrolyte platform for Organic redox flow batteries

Rohan Borah[a], Fraser R Hughson [b] and Thomas Nann *[a]

[a] Rohan Borah, Prof. Thomas Nann
School of Mathematical and Physical Sciences
University of Newcastle
Newcastle, New South Wales, 2308, Australia
E-mail: thomas.nann@newcastle.edu.au

[b] Fraser R. Hughson
School of Chemical and Physical Sciences
Victoria University of Wellington
Wellington, 6140, New Zealand

Supporting information for this article is given via a link at the end of the document.

**Abstract:** *Redox flow batteries (RFBs) have attained popularity as large-scale energy storage systems for wind and solar powered grids. State-of-the-art RFB systems are based on highly corrosive and/or flammable electrolytes. Organic redox active species for RFBs are gaining commercial traction, but there exists a compromise in choosing aqueous or non-aqueous electrolytes in terms of rate capabilities, energy density, safety, and cost. While modification of organic redox molecules to mitigate these issues is predominant in the literature, the search for novel electrolyte systems is scarce. We hereby present micro-emulsion-based electrolytes as an alternative for the next-generation organic RFBs. Micro-emulsion electrolytes (MEs) come with the benefits of decoupled solubility and ionic conductivity, wide electrochemical windows, non-flammability, simple modes of production and low costs of constituent chemicals. Electrochemical characteristics of organic redox species in MEs has been assessed to suit RFB requirements and a proof-of-concept flow cell with a widely studied redox system and a commercial membrane is exhibited. The compositions of MEs can be tuned to suit the redox system which renders them as a platform of electrolytes for all organic redox systems to consider and benefit from.*

With renewable energy becoming cheaper every day, the shift to a renewable economy depends largely on efficient and inexpensive storage of large volumes of energy.[1–4] Redox flow batteries decouple power and energy, offering flexibility in scales of application. The state-of-the-art Vanadium RFB has achieved breakthrough, but relies on a moderately abundant material, is limited in voltage by water-splitting and employs a highly corrosive acid electrolyte.[5–7] RFBs based on commercially available and inexpensive organic redox molecules (ROMs), can be a more elaborate alternative.[8] However, limited aqueous solubilities of the ROMs limit volumetric capacities. Organic solvents offer a wider range of ROM solubility but suffer low conductivity, high electrolyte costs and flammability risks, all of which compound on scale-up.[9] Research has been focused on molecular modification of ROMs, for aqueous solubility. Despite successful demonstrations of such systems, the thermodynamic limit of 1.23 V to an aqueous electrochemical cell pertains; not to mention the costs involved in scalable production of the modified ROMs.[10–13] Modified ROMs can change their redox behaviour, not always for the better. The fact that material utilisation is also largely affected by the solubility limits of charged states is a less scrutinized detail.[13] Conventional aqueous or non-aqueous electrolytes cannot alleviate all these challenges for high-energy density ORFBs.[14,15] However, electrolyte modification as opposed to ROM modification, can be industrially less onerous and more environmentally benign and yet an unpopular strategy. Alternate electrolytes for RFBs in the literature are limited to ionic liquids,[16–18] deep eutectics,[18–20] and liquid ROMs,[21] which involve issues primarily in costs, conductivity/viscosity and safety. As an alternative electrolyte for ORFBs, this research proposes the use of micro-emulsion (ME) based electrolytes for next generation organic RFBs (ORFBs). MEs have been suggested as possible ORFB electrolytes recently,[22] however, to our best knowledge a functional cell with an ME electrolyte is yet to be demonstrated. In validation, fundamental aspects of electrochemistry in micro-emulsions have been assessed and a proof-of-concept cell with a functional membrane has been demonstrated.

Micro-emulsions are thermodynamically stable dispersions of immiscible liquids stabilised by amphiphiles. MEs mostly exhibit characteristics similar to pure solutions, some differing aspects can be beneficial for ORFBs. While the resultant polarity of a mixed solvent is an intermediate of the individual polarities, MEs are micro-heterogeneous, thereby conserving the individual solvent polarities. The biphase offers decoupled ROM solubilities and ionic conductivities, two crucial but counteractive properties for ORFB electrolytes. This allows for electrochemistry of both polar and non-polar redox species in an electrolyte that offers aqueous conductivity.[23–26] An exhibition of this is seen in the cyclic voltammograms of Potassium ferrocyanide and Fullerene ($C_{60}$), in two homologous MEs of water/1-butanol/SDS/toluene merely by changing the o:w ratio in Fig.1a. It is to be noted that aqueous voltammetry of bare $C_{60}$ in a predominantly aqueous solution is unreported to our best knowledge. Not just pristine ROMs, but solubility change upon charging a ROM can also be accommodated in such a bi-phase, the composition of which can be tuned specific to a ROM and its charged state(s), a fact that is demonstrated in a proof-of-concept cell eventually.

**Table 1.** Compositions, electrochemical windows, and conductivities of representative blank micro-emulsions.

| ME | Composition [a] (o/w/s/cs, wt%) | Electrochemical Window (V) | Conductivity (mS/cm) [b] |
|----|---|---|---|
| 1 | 3.2/82.1/4.9/9.8 | 3.2 | 7.51 |
| 2 | 3.2/82.1/4.9/9.8 | 3.5 | 7.51 |
| 2h | 31.0/40.0/9.7/19.3 | 3.5 | 6.72 |
| 3 | 1.3/72.0/26.7/0.0 | 3.9 | 11.05 |
| 4 | 26.0/36.5/0.0/37.5 | 2.9 | 1.49 |

[a] ME 1-2h are based on oil/water/Sodium dodecyl sulphate (SDS)/butanol with 1 with cyclohexane as oil and 2,2h with toluene as oil. ME3 is based on toluene/water/Triton X-100 (TX-100). ME4 is a surfactant-free system of Dichloromethane/water/ethanol [b] Conductivities of ionic surfactant (SDS) based MEs is without added salt and the rest are with an overall KCl concentration of 0.1 M.

A concurrent benefit of using MEs as electrolytes is the extended electrochemical window. The micro-structure of the electrode-electrolyte interface (EEI) that governs the electrochemical window in MEs depend on the surface nature of the electrode and the surfactant, hence tuneable.[27] Fig.S1 shows the electrochemical windows of 5 representative MEs measured on a glassy carbon electrode at 100 mV/s and the compositions as listed in Table1. While all the MEs exhibit a substantially extended window compared to an aqueous KCl solution, the effect is more pronounced in case of the surfactant-based systems. On comparison to the aqueous solutions of the respective surfactants, the windows are at least 0.5 V higher for the MEs. This leads to the hypothesis that; on a hydrophobic carbonaceous electrode the electrode-electrolyte interface (EEI) is preferentially a layer of oil and surfactant. This hydrophobic layer suppresses reactions of water on the electrode by limiting diffusion. While the nature and amount of oil do not significantly affect the window, the nature of surfactant or rather it's hydrophile lipophile balance has major implication. TX-100 (HLB 13.5) based ME3 has a window 0f 3.9 V as opposed to SDS (HLB 40) MEs (<3.5). The aromatic head group of TX-100 leads to a more compact EEI than that of the aliphatic tail group in SDS, leading to increased overpotentials for water splitting. These results and the hypothesised EEI corroborates well to the findings from Neutron scattering analysis of the EEI of Tween based MEs on hydrophobic silane electrodes reported by Peng *et al*.[22]

Diffusion and kinetics of electron transfer of ROMs are important measures of electrolytes and affects rate capabilities of ORFBs. The two parameters for Ferrocene (Fc) in three representative MEs were analysed. Fig 2. a, b, c shows the CVs of 10 mM Fc in ME2, ME2h and ME3 at scan rates between 10-500 mV/s. While the low oil-content MEs, ME2 and ME3 show reversible oxidation with conventional peak separations, ME2b showed a resistively distorted voltammogram indicated by the large peak separations. This can be expected considering that ME2b comprises of 31 wt% toluene (to which no supporting electrolyte is added) making the EEI resistive. An interesting aspect of the CVs is the variance of peak currents densities ($j_p$) across the three MEs. Despite having the same overall concentration of 10 mM Fc, $j_p$ values are not only different but also do not correspond to the Fc concentrations in the toluene phase. This suggests that $j_p$ of a redox species in MEs is rather dictated by the diffusion. So, why is the diffusion of Fc different in three different MEs having the same dissolving phase; *i.e.* toluene? The apparent diffusion of a ROM in MEs governed by the composition dependant micro-structure and the partitioning of the redox solute in the bi-phase. For a predominantly oil-soluble species such as Fc, in case of the low-oil MEs, the peak current density is limited by the slower diffusion of the oil droplets from the bulk to the electrode arising from the o/w structure. Since the high-oil ME adopts a bi-continuous structure comprising of an oil-continuous phase, the apparent diffusion of Fc in the ME is comparable to the diffusion in the oil phase, thereby leading to higher current densities despite lower resultant concentration in the oil phase. Diffusion coefficients ($D_0$) and rate constants of electron transfer ($k_0$) were analysed from scan rate variation of cyclic voltammograms using the Nicholson-Lavagnini method. (Fig. S2-S4) The values are reported in Table 2.

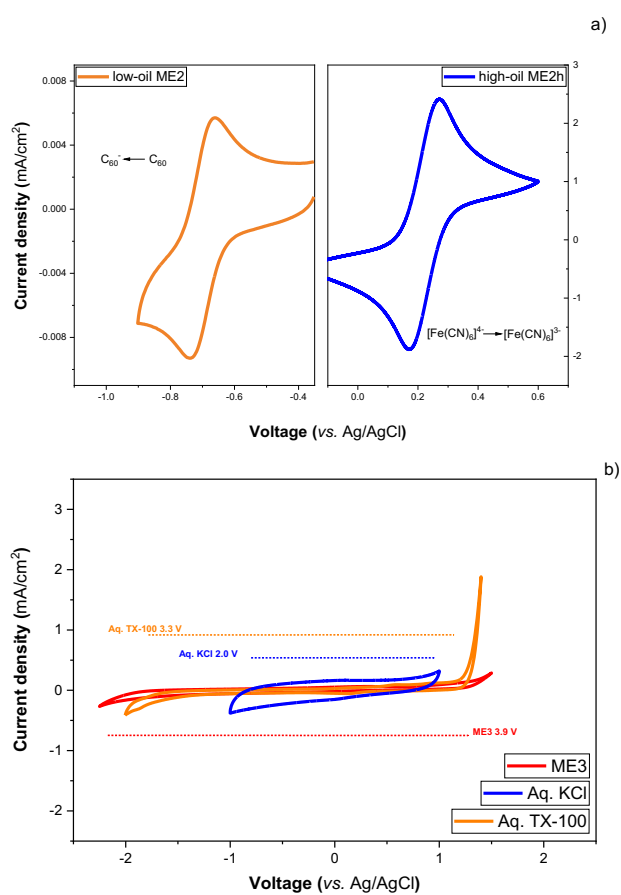

**Figure 1.** a) Cyclic voltammograms of (--) 1.4 mM $C_{60}$ in high-oil ME2h and (--) 10 mM $K_4[Fe(CN)_6]$ in low-oil ME2 on glassy carbon at 10 mV/s. b) Electrochemical window analysis of ME3 using cyclic voltammetry on a glassy carbon electrode at 100 mV/s.

McKay *et al* estimated the droplet-controlled diffusion for Fc to be of the order of $10^{-7}$ cm$^2$s$^{-1}$ based on Stokes-Einstein relation for typical o/w droplet sizes.[28] Both droplet MEs *viz.* ME2 and ME3 exhibit $D_0$ values in the suggested magnitude, while the five-fold lower $D_0$ for ME3 can be attributed to the higher viscosity arising from the TX-100 surfactant. Consistent with Mckay's findings, the $D_0$ of Fc in the bi-continuous ME2h is indeed an order of magnitude higher than the droplet MEs. However, the value is roughly an order of magnitude lower than in neat organic solvents owing to obstruction effects from opposite phases, surfactant interfaces as well as increased viscosity in MEs. The subsequent $k_0$ calculations reveal that despite the bi-continuous system being freely diffusing for Fc, electron transfer is only as rapid as in the droplet ME. This can be due to the resistive oil-rich EEI impeding the electron transfer process, considering that the oil phases in these MEs are effectively salt-free. Yet another interesting observation is that the $k_0$ in the TX-100 system is the highest despite the lowest $D_0$ value. The smaller hydrophobic part of TX-100 would lead to a less predominant EEI thereby making the electron transfer relatively facile. Thus, the electron transfer of oil soluble redox species is rapid in MEs irrespective of the complex EEI; corroborated by the $k_0$ values comparable to the VRFB redox reactions.[29] This entails that MEs as ORFB electrolytes are feasible as far as kinetics and rate capability is concerned.

Table 2. Diffusion coefficients and electron transfer rate constants for 10 mM Fc in three representative MEs.

| ME | D', Diffusion coefficient (cm$^2$.s$^{-1}$) | $k^{o'}$, Rate constant (cm.s$^{-1}$) |
|---|---|---|
| 2 | 3.74 x 10$^{-7}$ | 1.23 x 10$^{-3}$ |
| 2h | 3.53 x 10$^{-6}$ | 1.43 x 10$^{-3}$ |
| 3 | 5.35 x 10$^{-8}$ | 6.80 x 10$^{-3}$ |

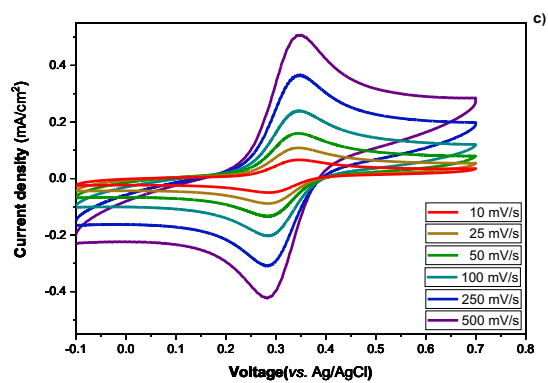

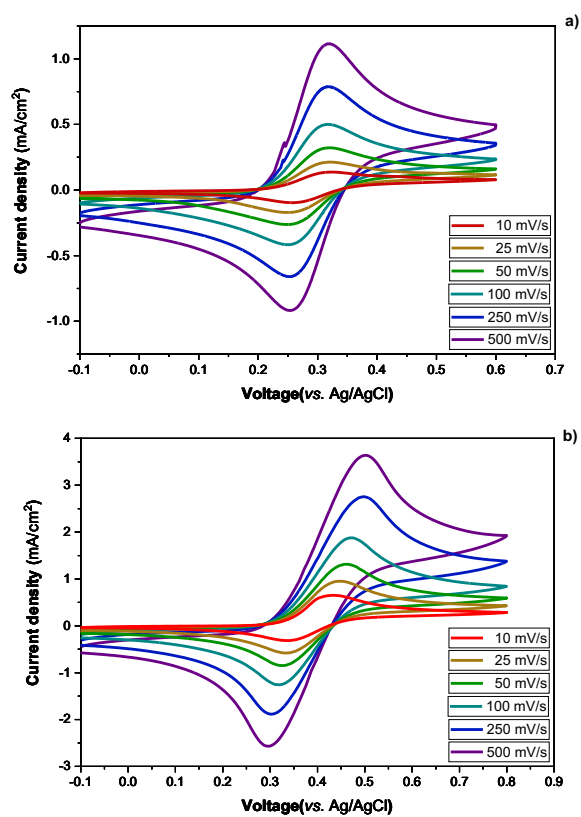

Figure 2. Cyclic voltammograms of 10 mM Fc in a) ME2 b) ME2h and c) ME3 on glassy carbon electrode between 10-500 mV/s.

The biphasic nature of MEs for the solubility of differently polar charged states is best expressed in the case of Methyl viologen dichloride, MVCl$_2$ (N, N'-Dimethyl-4,4'-bipyridinium dichloride) an extensively studied anolyte ROM for aqueous ORFBs.[6,8,11,30] MV$^{2+}$ undergoes two discreet 1 e$^-$ reductions to yield a radical cation (MV$^{+\cdot}$) and thereby the neutral species (MV$^0$). Insolubility of MV$^0$ in water as well as side reactions, makes the second reduction irreversible.[31] Fig. 3 shows the comparison between aqueous and ME3 voltammetry of 10 mM MVCl$_2$. The presence of the oil component allows for the dissolution of MV$^0$ rendering the second reduction reversible.

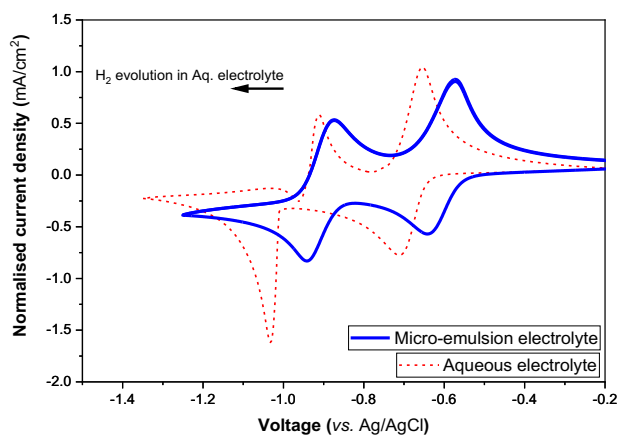

Figure 3. Cyclic voltammograms of MVCl$_2$ in ME3 electrolyte (——) and in aqueous electrolyte (----) on GC electrode at 10 mV/s.

N,N,N,2,2,6,6-Heptamethylpiperidinyloxy-4-ammonium chloride (TEMPTMA); a highly water-soluble catholyte ROM has frequently been coupled with MVCl$_2$ to develop high performing ORFBs which are limited in applicability mostly by solubility of the ROMs and cell voltage.[30] Substitution of the hydrophilic groups at *N*-position(s) in the viologen for better solubility or use of non-aqueous electrolytes have been reported but haven't been viable so far.[11] Considering the solubility of all viologen charged states, higher voltages composition based tunability of volumetric capacity and, MEs can potentially improve the system.

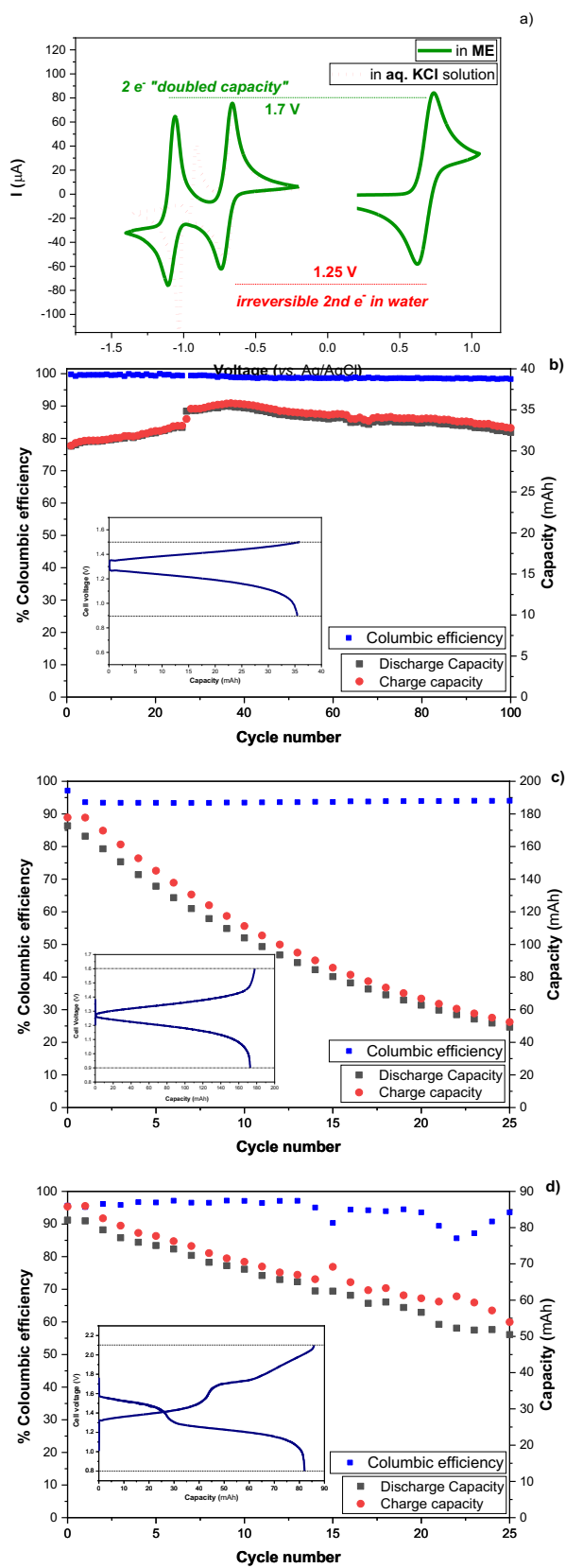

**Figure 4.** a) Comparison of theoretical voltages of a TEMPTMA/MVCl$_2$ in aqueous and ME electrolytes. 1 e$^-$ Galvanostatic charge-discharge for b) 0.1 M electrolyte, c) 0.5 M electrolyte and d) 2e$^-$ Galvanostatic charge-discharge for 0.1 M electrolyte. Insets: Typical Voltage vs. Capacity curves for each GCD.

A proof-of-concept ME based RFB was thereby implemented using 0.1 M active material electrolytes, a 5 cm$^2$ flow cell and two peristaltic pumps (20ml/min). Galvanostatic charge discharge (GCD) data at 20 mA/cm$^2$ for one-electron cycling at within voltage cut-offs of 1.5 and 0.9 V is shown in Fig4b. A columbic efficiency of 98.97% was achieved over the 100 cycles. The GCD results with the 0.5 M active material electrolyte, Fig 4c inset; showed typical charge-discharge plateaus maintaining a columbic efficiency of 93.8%. However, rapid capacity fade was observed during the cycling which was found to be caused by MV$^{2+}$ crossover in post-cycling voltammetry (Fig. S5). The fact that the capacity fade was much extensive over 25 cycles of 0.5 M electrolyte cycling compared to 100 cycles of 0.1 M electrolyte cycling implies that it is time dependant and not cycle dependant, thereby necessitating re-assessment of membrane-electrolyte compatibility. Nevertheless, the system was operated for 2e$^-$ cycling within voltage cut-off of 2.1 and 0.8 V, 20 mA/cm$^2$ current density as shown in Fig 4d. The two clearly defined charge-discharge plateaus were evident of a reversible 2 e$^-$ process. No gas evolution at either electrode was observed, implying the extended electrochemical window of MEs is achievable in a full-cell setup. Moreover, the reversibility of the 2$^{nd}$ electron cycling was significantly improved by increasing the oil content in the ME from 1.33 wt% to 10.7 wt% (Fig. S6) validating the tuneability of ME electrolytes for better charged state solubility. With the high aqueous solubility of MVCl$_2$ (2.5M) and the two mole equivalent electrons, volumetric capacities of up to 67 Ah/L can be achieved considering 50% oil content for reversible electrochemistry.

In summary, a new platform of electrolytes for ORFBs has been developed and its electrochemical characteristics validated. These micro-emulsion electrolytes are electrochemically stable, non-flammable, neutral pH and composed of inexpensive, industrially scaled chemicals. While these electrolytes can be tuned to suit ROMs with different solubilities, high operational voltages and energy densities can potentially be achieved without splitting water. With improvement in membrane-electrolyte compatibility and optimisation of electrolyte composition, MEs pose as an alternative to expensive, flammable, and toxic non-aqueous electrolytes.

## Acknowledgements

We acknowledge Global Connections fund, Department of Industry, Innovation and Science, Australia for financial support, Jena Batteries GmbH, Jena, Germany for providing the redox materials and Prof. Christina Roth, Chair of Electrochemical Process Engineering, University of Bayreuth, Germany for hosting the research during the COVID-19 pandemic.

**Keywords:** micro-emulsion • electrochemistry • electrolyte • redox flow battery • energy storage

[1] R. F. Service, *Science* **2019**, *365*, 108–108.
[2] G. He, J. Lin, F. Sifuentes, X. Liu, N. Abhyankar, A. Phadke, *Nat Commun* **2020**, *11*, 2486.
[3] M. Arbabzadeh, R. Sioshansi, J. X. Johnson, G. A. Keoleian, *Nat Commun* **2019**, *10*, 3413.
[4] B. Dunn, H. Kamath, J.-M. Tarascon, *Science* **2011**, *334*, 928–935.
[5] R. F. Service, *Science* **2018**, *362*, 508–509.
[6] P. Alotto, M. Guarnieri, F. Moro, *Renewable and Sustainable Energy Reviews* **2014**, *29*, 325–335.

[7] W. Wang, Q. Luo, B. Li, X. Wei, L. Li, Z. Yang, *Advanced Functional Materials* **2013**, *23*, 970–986.
[8] J. Winsberg, T. Hagemann, T. Janoschka, M. D. Hager, U. S. Schubert, *Angewandte Chemie International Edition* **2017**, *56*, 686–711.
[9] H. Chen, G. Cong, Y.-C. Lu, *Journal of Energy Chemistry* **2018**, *27*, 1304–1325.
[10] V. Singh, S. Kim, J. Kang, H. R. Byon, *Nano Res.* **2019**, *12*, 1988–2001.
[11] Y. Liu, Y. Li, P. Zuo, Q. Chen, G. Tang, P. Sun, Z. Yang, T. Xu, *ChemSusChem* **2020**, *13*, 2245–2249.
[12] J. Winsberg, C. Stolze, S. Muench, F. Liedl, M. D. Hager, U. S. Schubert, *ACS Energy Letters* **2016**, *1*, 976–980.
[13] N. H. Attanayake, J. A. Kowalski, K. V. Greco, M. D. Casselman, J. D. Milshtein, S. J. Chapman, S. R. Parkin, F. R. Brushett, S. A. Odom, *Chem. Mater.* **2019**, *31*, 4353–4363.
[14] W. Liu, W. Lu, H. Zhang, X. Li, *Chemistry – A European Journal* **2019**, *25*, 1649–1664.
[15] Z. Tang, A. P. Kaur, A. Pezeshki, S. Modekrutti, F. Delnick, T. Zawodzinski, G. Veith, S. Odom, **2021**, DOI 10.26434/chemrxiv.13739692.v1.
[16] A. Ejigu, P. A. Greatorex-Davies, D. A. Walsh, *Electrochemistry Communications* **2015**, *54*, 55–59.
[17] R. Chen, R. Hempelmann, *Electrochemistry Communications* **2016**, *70*, 56–59.
[18] M. H. Chakrabarti, F. S. Mjalli, I. M. AlNashef, Mohd. A. Hashim, Mohd. A. Hussain, L. Bahadori, C. T. J. Low, *Renewable and Sustainable Energy Reviews* **2014**, *30*, 254–270.
[19] G. Cong, Y.-C. Lu, *Chem* **2018**, *4*, 2732–2734.
[20] N. S. Sinclair, D. Poe, R. F. Savinell, E. J. Maginn, J. S. Wainright, *J. Electrochem. Soc.* **2021**, *168*, 020527.
[21] J. Huang, L. Cheng, R. S. Assary, P. Wang, Z. Xue, A. K. Burrell, L. A. Curtiss, L. Zhang, *Advanced Energy Materials* **2015**, *5*, 1401782.
[22] J. Peng, N. M. Cantillo, K. M. Nelms, L. Roberts, G. A. Goenaga, A. E. Imel, B. Barth, M. D. Dadmun, L. Heroux, D. G. Hayes, T. A. Zawodzinski, *ACS Appl. Mater. Interfaces* **2020**, DOI 10.1021/acsami.0c07028.
[23] M. Kunitake, E. Kuraya, D. Kato, O. Niwa, T. Nishimi, *Current Opinion in Colloid & Interface Science* **2016**, *25*, 13–26.
[24] F. M. Menger, A. R. Elrington, *J. Am. Chem. Soc.* **1991**, *113*, 9621–9624.
[25] R. A. Mackay, S. A. Myers, Liakatali. Bodalbhai, Anna. Brajter-Toth, *Anal. Chem.* **1990**, *62*, 1084–1090.
[26] M. O. Iwunze, Artur. Sucheta, J. F. Rusling, *Anal. Chem.* **1990**, *62*, 644–649.
[27] "A 2.7 V Aqueous Supercapacitor Using a Microemulsion Electrolyte** - Hughson - - Batteries & Supercaps - Wiley Online Library," can be found under https://chemistry-europe.onlinelibrary.wiley.com/doi/full/10.1002/batt.202000314, **n.d.**
[28] R. A. Mackay, S. A. Myers, A. Brajter-Toth, *Electroanalysis* **1996**, *8*, 759–764.
[29] N. Roznyatovskaya, J. Noack, K. Pinkwart, J. Tübke, *Current Opinion in Electrochemistry* **2020**, *19*, 42–48.
[30] T. Janoschka, N. Martin, M. D. Hager, U. S. Schubert, *Angewandte Chemie International Edition* **2016**, *55*, 14427–14430.
[31] C. L. Bird, A. T. Kuhn, *Chem. Soc. Rev.* **1981**, *10*, 49–82.

## Entry for the Table of Contents

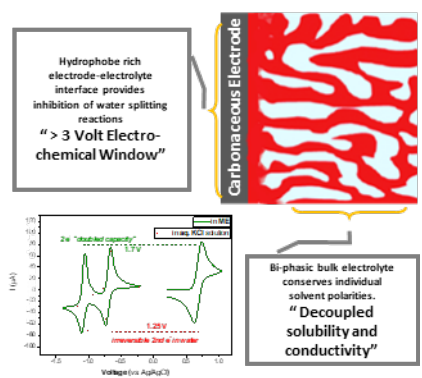

This study assesses electrochemistry in micro-emulsion electrolytes unravelling specific benefits for organic redox flow batteries ranging from an extended voltage window to rapid electron transfer kinetics from both aqueous and non-aqueous phases. A proof-of-concept redox flow cell with a commercial membrane is demonstrated introducing micro-emulsions as platform of electrolytes that can be cost effective, neutral pH, non-flammable alternatives to conventional RFB electrolytes.

Supporting Information

# Re-discovering micro-emulsion electrolytes: a biphasic electrolyte platform for Organic redox flow batteries

Rohan Borah, Fraser R Hughson and Thomas Nann*

**Abstract:** *Redox flow batteries (RFBs) have attained popularity as large-scale energy storage systems for wind and solar powered grids. State-of-the-art RFB systems are based on highly corrosive and/or flammable electrolytes. Organic redox active species for RFBs are gaining commercial traction, but there exists a compromise in choosing aqueous or non-aqueous electrolytes in terms of rate capabilities, energy density, safety, and cost. While modification of organic redox molecules to mitigate these issues is predominant in the literature, the search for novel electrolyte systems is scarce. We hereby present micro-emulsion-based electrolytes as an alternative for the next-generation organic RFBs. Micro-emulsion electrolytes (MEs) come with the benefits of decoupled solubility and ionic conductivity, wide electrochemical windows, non-flammability, simple modes of production and low costs of constituent chemicals. Electrochemical characteristics of organic redox species in MEs has been assessed to suit RFB requirements and a proof-of-concept flow cell with a widely studied redox system and a commercial membrane is exhibited. The compositions of MEs can be tuned to suit the redox system which renders them as a platform of electrolytes for all organic redox systems to consider and benefit from.*

**Table of Contents**

## Experimental Procedures

**Chemicals and instruments.** All chemicals were received from Sigma Aldrich and used without further treatment. N, N'-Dimethyl-4,4'-bipyridinium dichloride ($MVCl_2$) and N,N,N,2,2,6,6-Heptamethylpiperidinyloxy-4-ammonium chloride (TEMPTMA) were recieved as aqueous solutions of 45 wt% and 50 wt% strength, courtesy of Jena batteries GmbH, Germany. All electrochemical measurements were done on a 16 channel VMP-3 potentiostat/galvanostat (Biologic). Conductivity measurements were done on ECtstr11 conductivity meter (Eutech Instruments).

**Sample preparation.** The micro-emulsions were prepared by accurately weighing the components according to their respective wt% and mixing in the following manner. In a stoppered erlenmeyer flask, to a slurry of the surfactant/cosurfactant, the oil phase was mixed using magnetic stirring. Subsequently, the aqueous phase was added dropwise under magnetic stirring. The ROMs were dissolved in the respective phase before ME preparation. A clear, homogenous ME was finally obtained by ultrasonication. Potassium chloride was added as supporting electrolyte at the end, to applicable MEs.

**Electrochemical methods.** Cyclic voltammetry analyses were done on a 3 mm diameter glassy cabon disc working electrode with PEEK shrouding (Metrohm), a platinum rod counter electrode and a Ag/AgCl reference electrode with sat. KCl electrolyte. The GC electrode was thoroughly polished with 0.3 micron and 0.05 micron Alumina slurries (Buehler) before each measurement. All samples were purged with $N_2$ prior to analysis to displace dissolved $O_2$. For electrochemical window analysis, CVs were performed within the voltage bounds of the decomposing reactions and an arbitrary current density cut-off of 0.35 mA/cm$^2$ was used to assess the voltage stability window. For diffusion coefficient and rate constant measurements, the third cycle of every cyclic voltammogram was used for analysis such that an equilibriated system was considered. All CVs were acquired wioth 85% ZIR correction.

Diffusion coefficients were measured using the Randles-Sevcik equation;

$$i_p = 0.4463nFAC(nF\upsilon D/RT)^{0.5}$$

Where, $i_p$ is the maximum current in Amps. n is the number of transferred electrons *viz.* 1. F is the faraday constant, 96485.33 C.mol$^{-1}$. A is the electrode area 0.07065 cm$^2$. C is the bulk analyte concentration in mol.cm$^{-3}$. $\upsilon$ is scan rate in V/s. R is the universal gas constant 8.3144 J.K$^{-1}$.mol$^{-1}$. T is temperature, 298.15 K. Diffusion coefficients were obtained in cm$^2$s$^{-1}$. $i_p$ vs. $\upsilon^{0.5}$ was plotted, the slope of which gave the value of D. The equation was applied separately for the cathodic and anodic peak currents to obtain two diffusion coefficients, the average of which was used for subsequent calculations.

Rate constants were measured using the Nicolson-Lavagnini equation which defines a kinetic parameter ψ as per the following equations;

$$\Psi = (-0.6288 + 0.0021 \Delta E_p)/(1-0.017 \Delta E_p)$$

$$\Psi = k^0 (\pi DnF/RT)^{-0.5}\upsilon^{-0.5}$$

where $\Delta E_p$ is the peak separation between the cathodic and anodic peaks in the CV, in mV. $k^0$ was obtained in cm.s$^{-1}$ from the slope of the linear plot of Ψ vs. $\upsilon^{-0.5}$.

Galvanostatic charge-discharge measurements were conducted using a 5 cm$^2$ flow cell comprising of aluminium end plates, PEEK flow channels and graphite current collectors, sealed using EPDM rubber seals. Graphite felt electrodes (GFA, 4.6 mm) purchased from SGL were soaked in the respective electrolyte for 1 h prior to cell assembly. The commercial anion exchange membrane was pretreated in 1 M KCl for 24-72 hours prior to assembly, the solution being frequently changed. The electrolytes used for 1e$^-$ cycling were 0.1 M active material and 0.5 M active material for the low capacity and high capacity cycling respectively. For the 2 e$^-$ cycling, 0.2 M TEMPTMA and 0.1 M $MVCl_2$ concentrations were used. All electrolytes had a net volume of 30 ml and comprised of 0.5 M KCl as supporting salt. The electrolytes were pumped using Simdos 10 diaphragm pump (KNF) with a PTFE head and Tygon E3603 (masterflex) at 20 ml.min$^{-1}$. A constant charging and discharging current of ±100 mA (20 mA.cm$^{-2}$) was applied within the voltage cut-offs as specified in the main text.

## Results and Discussion

**Electrochemical windows of MEs.**

**Figure S1.** Cyclic voltammograms of 5 representative MEs compared to 0.1 M Aq. KCl and respective aqueous surfactant solutions. On glassy carbon working electrode at 100 mV/s. Electrochemical windows were considered using a 0.35 mA/cm$^2$ cut-off to current density.

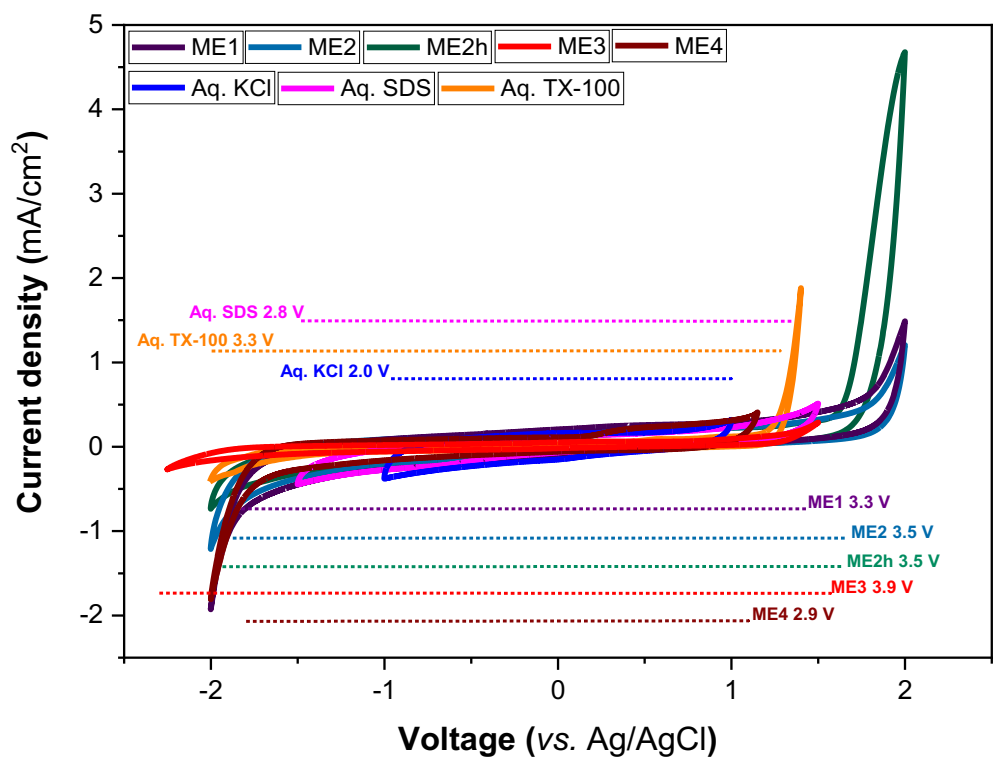

## Diffusion and electrode kinetics in MEs.

**Figure S2.** a) Randles-Sevcik plot ($i_p$ vs. $\upsilon^{0.5}$) for oxidation and subsequent reduction and corresponding b) Nicholson-Lavagnini plot of 10 mM Ferrocene in **ME2** on glassy carbon.

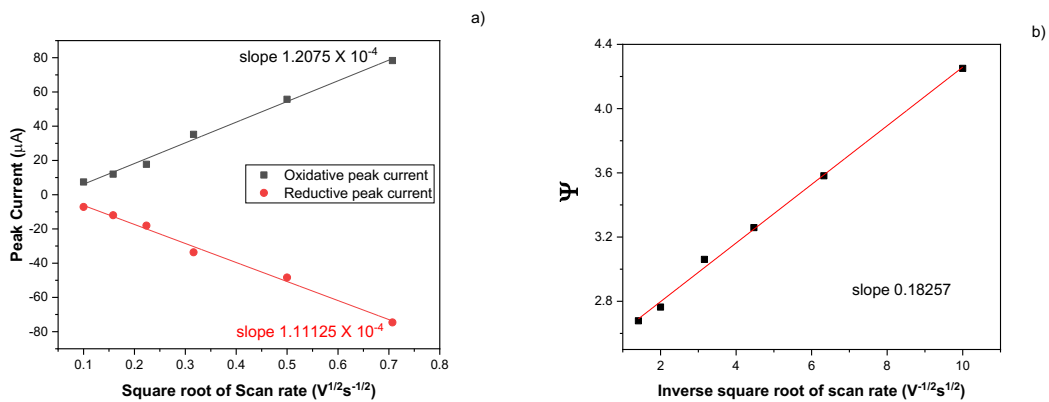

**Figure S3.** a) Randles-Sevcik plot ($i_p$ vs. $\upsilon^{0.5}$) for oxidation and subsequent reduction and corresponding b) Nicholson-Lavagnini plot of 10 mM Ferrocene in **ME2h** on glassy carbon.

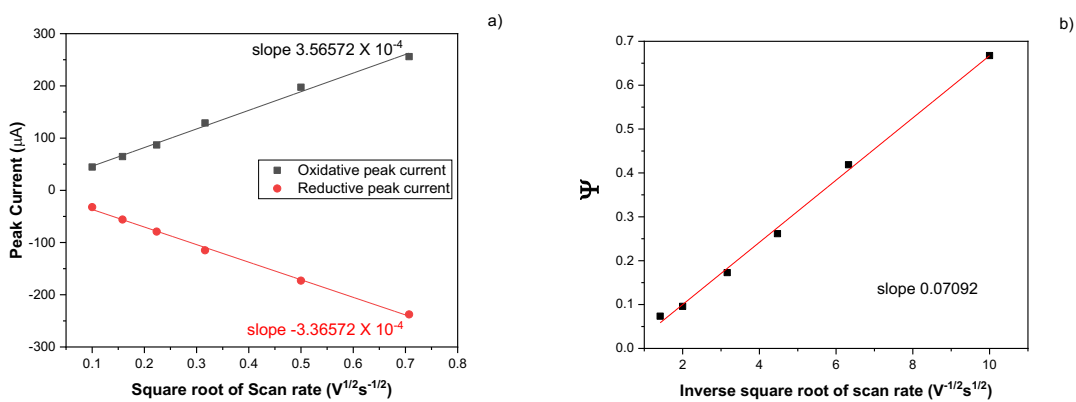

**Figure S4.** a) Randles-Sevcik plot ($i_p$ vs. $\upsilon^{0.5}$) for oxidation and subsequent reduction and corresponding b) Nicholson-Lavagnini plot of 10 mM Ferrocene in **ME3** on glassy carbon.

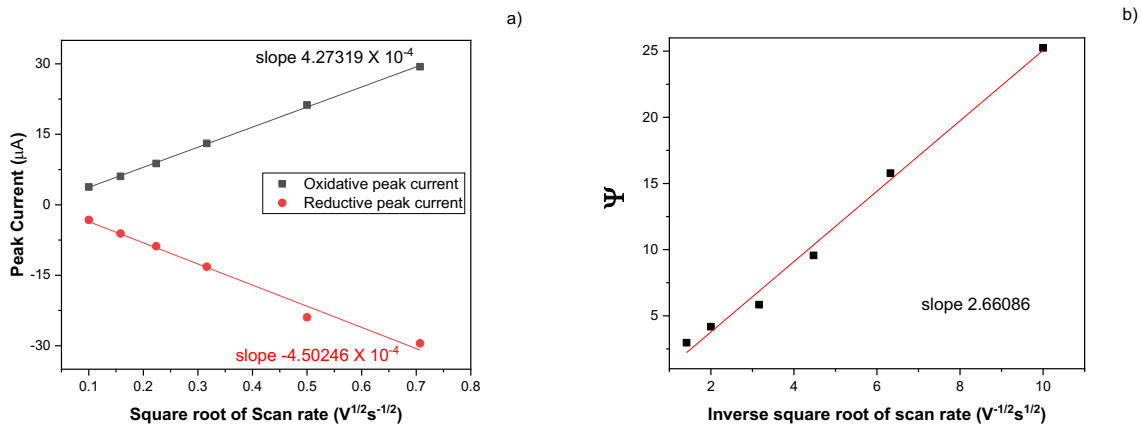

**Figure S5.** Cycling voltammogram of 0.1 M Anolyte after Galvanostatic charge discharge. The signal from cross-over of TEMPTMA catholyte species is highlighted.

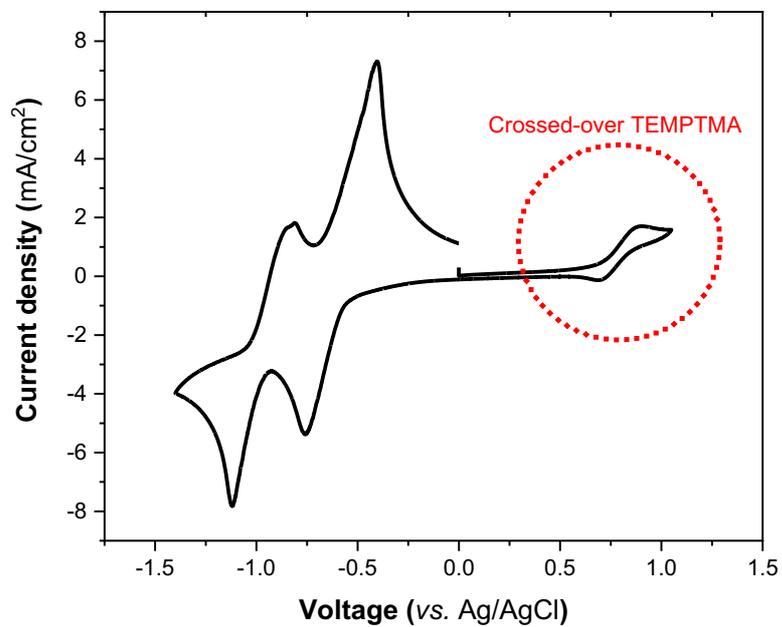

**Figure S6.** Comparison of Galvanostatic charge-discharge data of low and high oil ME shows the effect of increasing oil content from 1.33% (----) to 10.3% (----) on the reversibility of the second charge-discharge plateau.

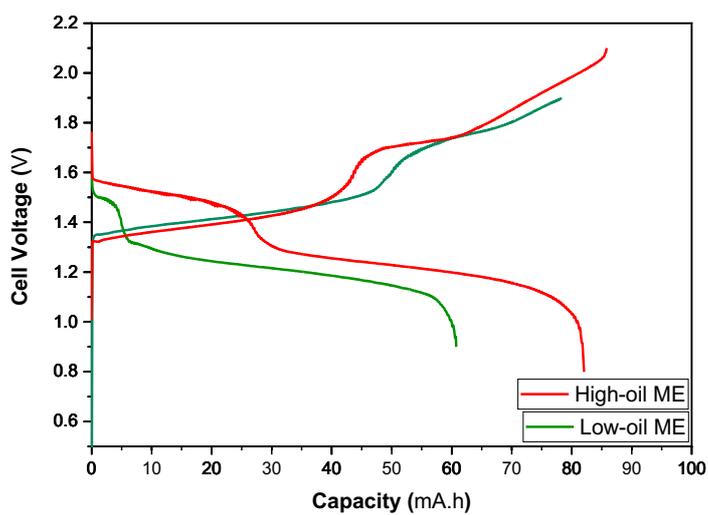

## Author Contributions

Rohan Borah; Lead role in data curation, investigation, formal analysis, validation and writing of original draft

Fraser R Hughson; Supporting role in data curation, formal analysis, validation and writing of original draft

Thomas Nann; Lead role in funding acquisition and project administration. Supporting role in validation and writing of original draft.